\documentclass[lettersize,journal]{IEEEtran}
\usepackage{amsmath,amsfonts}
\usepackage{algorithmic}
\usepackage{algorithm}
\usepackage{array}
\usepackage[caption=false,font=normalsize,labelfont=sf,textfont=sf]{subfig}
\usepackage{textcomp}
\usepackage{stfloats}
\usepackage{url}
\newcommand{\bigO}{\mathcal{O}}
\usepackage{multirow}
\usepackage{enumitem}
\usepackage{verbatim}
\usepackage{graphicx}
\usepackage{cite}
\usepackage{xcolor,color,colortbl}
\usepackage{scalerel}
\usepackage{tikz}
\usetikzlibrary{svg.path}

\definecolor{orcidlogocol}{HTML}{A6CE39}
\tikzset{
  orcidlogo/.pic={
    \fill[orcidlogocol] svg{M256,128c0,70.7-57.3,128-128,128C57.3,256,0,198.7,0,128C0,57.3,57.3,0,128,0C198.7,0,256,57.3,256,128z};
    \fill[white] svg{M86.3,186.2H70.9V79.1h15.4v48.4V186.2z}
                 svg{M108.9,79.1h41.6c39.6,0,57,28.3,57,53.6c0,27.5-21.5,53.6-56.8,53.6h-41.8V79.1z M124.3,172.4h24.5c34.9,0,42.9-26.5,42.9-39.7c0-21.5-13.7-39.7-43.7-39.7h-23.7V172.4z}
                 svg{M88.7,56.8c0,5.5-4.5,10.1-10.1,10.1c-5.6,0-10.1-4.6-10.1-10.1c0-5.6,4.5-10.1,10.1-10.1C84.2,46.7,88.7,51.3,88.7,56.8z};
  }
}

\newcommand\orcidicon[1]{\href{https://orcid.org/#1}{\mbox{\scalerel*{
\begin{tikzpicture}[yscale=-1,transform shape]
\pic{orcidlogo};
\end{tikzpicture}
}{|}}}}

\usepackage{hyperref} 
\begin{document}

\title{Identification of Distorted RF Components via Deep Multi-Task Learning}

\author{Mehmet Ali~Ayg\"{u}l \orcidicon{0000-0002-1797-8238}\,,~\IEEEmembership{Student Member,~IEEE,}
Ebubekir Memi\c{s}o\u{g}lu \orcidicon{0000-0001-5137-8511}\,, Hakan Ali Çırpan \orcidicon{0000-0002-3591-6567}\,,~\IEEEmembership{Member,~IEEE,}
and~H\"{u}seyin~Arslan \orcidicon{0000-0001-9474-7372}\,,~\IEEEmembership{Fellow,~IEEE}

\thanks{This work has been submitted to the IEEE for possible publication. Copyright may be transferred without notice, after which this version may no longer be accessible.}
\thanks{This work was supported in part by the Scientific and Technological Research Council of Turkey (TÜBİTAK) under Grant No. 5200030 with the cooperation of Vestel and Istanbul Medipol University.}
\thanks{M. A.~Ayg\"{u}l and H. A.~Çırpan are with the Department of Electronics and Communications Engineering, Istanbul Technical University, Istanbul, 34467, Turkey. M. A.~Ayg\"{u}l is also with the Department of Research \& Development, Vestel, Manisa, 45030, Turkey (e-mails: mehmetali.aygul@ieee.org and cirpanh@itu.edu.tr).
}
\thanks{E.~Memi\c{s}o\u{g}lu and H.~Arslan are with the Department of Electrical and Electronics Engineering, Istanbul Medipol University, Istanbul, 34810, Turkey. (e-mails: ebubekir.memisoglu@std.medipol.edu.tr and huseyinarslan@medipol.edu.tr).}}


\maketitle

\begin{abstract}
High-quality radio frequency (RF) components are imperative for efficient wireless communication. However, these components can degrade over time and need to be identified so that either they can be replaced or their effects can be compensated. The identification of these components can be done through observation and analysis of constellation diagrams. However, in the presence of multiple distortions, it is very challenging to isolate and identify the RF components responsible for the degradation. This paper highlights the difficulties of distorted RF components' identification and their importance. Furthermore, a deep multi-task learning algorithm is proposed to identify the distorted components in the challenging scenario. Extensive simulations show that the proposed algorithm can automatically detect multiple distorted RF components with high accuracy in different scenarios.
\end{abstract}

\vspace{-0.2cm}


\vspace{-0.3cm}

\section{Introduction}
 
\par The radio frequency (RF) components in any wireless communication system can degrade over time due to overuse, misuse, or other environmental factors. Therefore, their reliable maintenance and testing are imperative to ensure the system's smooth operation. A critical part of this maintenance is the identification of any distorted RF components since they can lead to RF distortions in the signal \cite{angrisani2005error}. Generally, constellation diagrams are helpful in identifying the distorted components \cite{report}. However, in the presence of multiple distorted components, this approach fails since the constellation diagram becomes too cloudy for reliable identification of the distorted components. Moreover, new enablers of future wireless communication networks such as higher frequencies, higher modulation orders, massive multiple input and multiple output technologies are expected to be more vulnerable to RF impairments, which makes the problem of distorted components identification even more challenging \cite{mohammadian2021rf}.

\par In recent years, machine learning (ML), particularly deep learning (DL) has found its applications in complex problems pertaining to wireless communications such as channel estimation, signal detection, and radio resource allocation \cite{eldar2022machine}. Also, since distorted RF components identification is complex and ultimately a classification problem, it aligns well with the DL methodology. In literature, DL is used to estimate channel, phase noise (PN), and in-phase component of the signal $(I)$ and the quadrature component of the signal $(Q)$ gain imbalance (GI) in \cite{mohammadian2021deep}, and $I/Q$ offset, quadrature offset (QO), and, PN are jointly estimated in \cite{aygul2022joint}. Also, there has been increased interest in multi-task learning (MTL) in wireless communication research to address different learning-based problems simultaneously by using the relation between related problems \cite{wong2022transfer}. This approach aligns with the problem of distorted RF components identification in the presence of multiple impairments.

\par In this paper, firstly, we highlight the problem of identifying multiple distorted RF components. Then, we propose an algorithm to identify the distorted components using a deep MTL algorithm since it has the ability to learn multiple problems simultaneously. The algorithm is trained with generated distorted signals and their hardware's distortion statuses as the input and output datasets, respectively. To demonstrate the efficiency of the proposed algorithm, extensive simulation results are given in several scenarios. Accuracy, precision, recall, and $F_1$-score performance metrics are used for this validation.

\vspace{-0.3cm}
\section{The Model of RF Impairments}
\label{Section2}

\par The RF components are distorted over time, and they create RF impairments. The models of these impairments are explained in this part based on \cite{arslan}. Also, the effects of the RF impairments and their corresponding components are illustrated in Fig.~\ref{hardware_impairments}. In this figure, each hardware component is matched with different RF impairment characteristics. These illustrated impairments and their parameters are made for the single-carrier waveform. These impairments are $I/Q$ GI, QO, PN, frequency offset (FO), and $I/Q$ offset, where their parameters are denoted by $I_g$/$Q_g$, $\psi$, $\phi$, $f_o$, and $I_o$/$Q_o$, respectively. In Fig.~\ref{hardware_impairments}, parameters of the RF impairments are set as follows; $I_g=0.65$, $Q_g=0.42$, $\psi=60^{\circ} $, $\phi=68^{\circ} $, $f_o=42$ Hz, $I_o=-0.32$, and $Q_o=0.45$.

\subsubsection{$I/Q$ GI} After the signal is generated in the digital domain, it is converted to the analog domain by a digital to analog converter (DAC). Then, the signal is passed through a band-pass filter. If the filter is distorted, it makes the $I$ to be unequal to $Q$. Therefore, an imbalanced change in the original constellation is observed. The imbalance of $I/Q$ gain can be calculated as $(I_g/Q_g -1)\times 100$ in percent.

\subsubsection{QO} A phase offset of the sine and cosine signal generation causes QO in the phase shifter (PS). This impairment distorts the orthogonality between $I$ and $Q$ branches. Thus, the angle between $I$ and $Q$ branches is not 90$^{\circ}$. The signal with QO can be modeled as
\begin{equation}
x_{qo} (t) = \cos{(2\pi f_ct)}I_g\Re\{x(n)\}+ \sin{(2\pi f_ct+\psi)}Q_g\Im\{x(n)\},
\end{equation}
\noindent where $t$, $x(n)$, and $f_c$ signify time, $n$-th sample of the modulated time-domain signal, and carrier frequency respectively. $\Re$ and $\Im$ denote the real and imaginary parts of the signal, respectively.

\subsubsection{PN} 

\par A practical local oscillator (LO) causes PN by spreading at the desired spectrum. It is assumed that the PN is constant over the data. Therefore, the effect of the PN is the phase rotation of the symbols. A signal with PN can be modeled as 
\begin{equation}
x_{pn} (t) = x_{qo} (t)e^{j\phi},
\end{equation}
\noindent where $\phi$ is rotation of the phase.

\subsubsection{FO} The LO mismatch between the transmitter and receiver causes FO. It is modeled with a signal as
\begin{equation}
x_{fo} (t) = x_{pn}(t) e^{j2\pi f_o t}.
\end{equation}

\par FO causes a phase rotation at the constellation diagram. The amount of rotation introduced to each constellation point is dependent on the sample index and FO value, and it keeps growing. The constellation diagram becomes a circular shape due to the growing phase rotations.

\subsubsection{$I/Q$ offset} Mixers cause $I/Q$ offset impairment. Its effect is a shift at the constellation diagram, and it can be modeled with a signal as follows.
\begin{equation}
x_{iqo} (t) = x_{fo} (t) + I_{o} + jQ_{o},
\end{equation}
\noindent where $I_{o}$ and $Q_{o}$ represent $I$ and $Q$ offsets, respectively.

\par
\begin{figure}[!t]
\centering
\resizebox{.99\columnwidth}{!}{
\includegraphics[width=14cm]{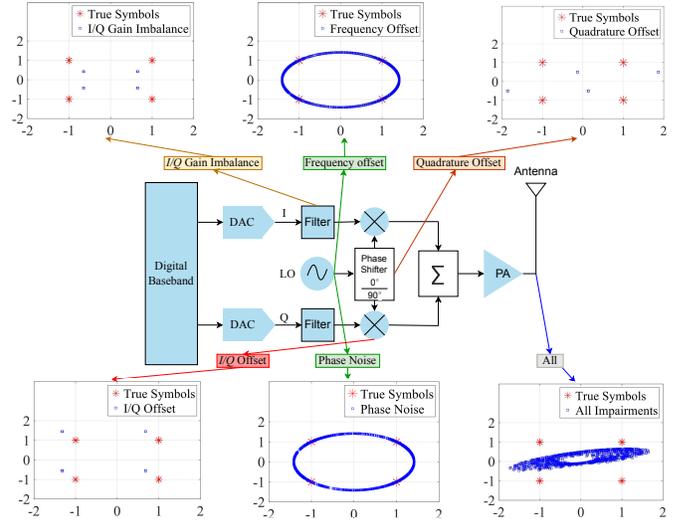}}
\vspace{-0.3cm}
\caption{A typical transmitter with various sources of RF impairments.}
\label{hardware_impairments}
\end{figure}

 When all aforementioned impairments affect the signal, the modulated symbols become very distorted, as plotted in Fig.~\ref{hardware_impairments} (All). The distorted signal causes the constellation diagram to be cloudy, which makes the identification problem more complex.

\par ML algorithms are promising for solving complex problems, especially when it is used with multiple hidden layers (DL algorithms). This is because multiple hidden layers can magnify the intrinsic distinctive features of the data while suppressing the irrelevant information at each layer.

\vspace{-0.3 cm}

\section{Identification of Distorted RF Components}

\par As shown in (4), the distorted RF components cause additive and multiplicative impairments in wireless communication systems. If only one or a few components are distorted in the system, distorted components can be identified by looking at the constellation diagram. However, multiple distorted RF components are considered in this paper, as shown in Fig.~\ref{hardware_impairments}. Therefore, the identification of distorted hardware components becomes a challenging problem. For example, let's assume that filter ($I/Q$ GI), PS (QO), and mixer ($I/Q$ offset) are distorted, then the distorted signal can be written as
\begin{subequations}\label{eqn}
\begin{align*}
r(t) = &\cos{(2\pi f_ct)}I_g\Re\{x(n)\}+ \sin{(2\pi f_ct+\psi)}Q_g\Im\{x(n)\} \\
   &+ I_{o} + jQ_{o}. \tag{\ref{eqn}} 
\end{align*}
\end{subequations}
\par (5) indicates that when the identification of the PS distortion is made from $r(t)$, the accuracy of the identification depends on QO, $I/Q$ GI, and $I/Q$ offset distortions, not only QO. Furthermore, when there are more distorted RF components that cause RF impairments, the problem becomes more complex, like a black box, and accurate identification of individual hardware component distortion becomes infeasible. Trial and error methods can be used to identify the distorted components. In such an approach, the components are replaced while monitoring the constellation diagram. However, this approach is not scalable since it requires checking after replacing each component.

\par DL-based algorithms are used for black box problems. However, the optimizations in DL are still based on a specific metric such as mean square error and classification accuracy. A single model or an ensemble of models is usually trained to solve the single problem for the optimization. Although acceptable performance is achieved by being laser-focused on a single problem, the performance of the original problem can be increased by trying to find optimum solutions for different problems simultaneously \cite{ruder2017overview}. This feature can also be applied to the problem of joint distorted RF components identification.

\par MTL is able to interpret useful information available in multiple problems to improve the identification performance of individual problems \cite{zhang2021survey}. Thus, it is promising to identify multiple distorted hardware components with MTL. This is especially the case where MTL has several hidden layers, referred to as deep MTL. Accordingly, this paper designs a deep MTL algorithm to identify multiple distorted hardware components jointly. There are two parts of the proposed algorithm. These parts are training and testing. The training part involves dataset generation, configuration, and DL training. Afterward, the trained deep MTL algorithm is used to identify the distorted RF components for the testing part.

\begin{figure}[!t]
\centering
\resizebox{0.95\columnwidth}{!}{
\includegraphics[width=14cm]{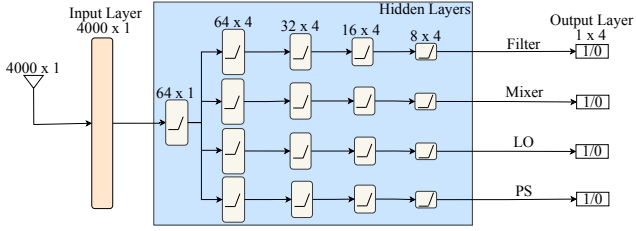}}
\vspace{-0.3cm}
\caption{The tuned deep MTL algorithm.}
\label{the_proposed_DL}
\end{figure}

\par In the training part, firstly, a signal to be transmitted is generated. Note that this signal is generated only once for all samples, including training, testing, and validation datasets to make the DL easy to learn the model\footnote{If the signal is generated for each training data sample, the DL algorithm needs to learn generated signal as well since the effect of impairments is different for each sample. This will increase the time complexity since more epochs and data samples will be required to learn the signals' characteristics.}. Afterward, the generated signal is passed through the RF components so that impairments of the RF components are added to the signal. Then, the distorted signal is obtained at the transmitter antenna and stored as input in vector format. Correspondingly, the distorted hardware components are indicated with ``1" and non-distorted components are indicated with ``0", and stored as the output. This process is repeated until a sufficiently large dataset is obtained. Based on the requirements of system complexity, performance, and memory, the dataset size can be selected. All of the hyperparameters of the deep MTL algorithm (e.g., the number of hidden layers, units in each hidden layer, and epoch) are optimized empirically according to the generalization capability and performance of the proposed algorithm. For this purpose, validation dataset\footnote{In the ML context, the validation dataset is used to provide an unbiased evaluation of a algorithm fit on the training dataset.} is used. More specifically, the hyperparameters are adjusted in such a way that the training and validation accuracies converge after some epochs.

\par The tuned deep MTL algorithm is demonstrated in Fig.~\ref{the_proposed_DL}. The algorithm consists of an input layer, five fully connected hidden layers, and an output layer. For the input layer, 4000 units are used. Afterward, joint identification of the distorted hardware components is performed in the first hidden layer with 64 units. Next, the characteristics of each distorted RF component is learned in the four hidden layers with the 64, 32, 16, and 8 units, respectively. Finally, all of the hardware component distortions are identified in the output layer with a unit per hardware component. The activation function with a rectified linear unit is used in all layers. The number of total parameters of the tuned deep MTL algorithm is 283716 parameters. Moreover, the algorithm is trained with a batch size of 16 and 10000 epochs. The logarithmic loss function for binary classification and efficient adaptive moment estimation for adaptive learning rate optimization with 0.000001 are used during the training part. After the training part, the testing part that characterizes the run-time operation of the proposed algorithm starts.

\par
\begin{figure}[!t]
\centering
\resizebox{0.99\columnwidth}{!}{
\includegraphics[width=14cm]{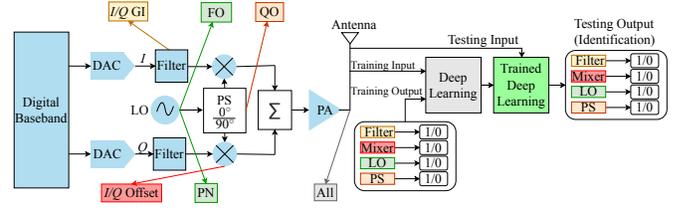}}
\caption{Identification of the distorted hardware components.}
\label{the_proposed_algorithm}
\end{figure}

 In the testing part, firstly, a testing signal distorted by multiple RF components is generated. Afterward, this distorted signal is fed to the trained DL algorithm. Then, the trained DL algorithm identifies the distorted hardware components according to the testing input (distorted testing signal). The overall steps of the proposed algorithm are illustrated in Fig.~\ref{the_proposed_algorithm}.

\par The computational complexity of the proposed algorithm mainly depends on the learning complexity of the training and testing parts. This work adopts a deep MTL algorithm. This algorithm includes an input layer, five hidden layers, and an output layer. The algorithm has $a$ units in the input layer. In the hidden layers, $b$ hidden units exist for joint learning. Then, four more hidden layers are used which have $c$, $d$, $e$, and $f$ hidden units, respectively. Finally, $g$ output units are used for each impairment. Therefore, the overall training computational complexity of this algorithm is $\bigO(ps\times (ab+4(bc+cd+de+ef+fg)))$, where the numbers of epochs and training examples are signified by $p$ and $s$, respectively. Moreover, the number of trials to select the optimum hyperparameters and the computational complexity of the validation part can be added to the training complexity. It is noteworthy that trials and the numbers of validation data depend on the reliability and complexity requirements of the application. Moreover, the computational complexity of testing samples is roughly equal to half of the training samples \cite{he2015convolutional}.

\vspace{-0.2cm}
\section{Simulation Results}

\par In the simulations\footnote{Simulation code is provided to reproduce the dataset in this paper:
https://github.com/mehmetaaygul/Hardware-Problem-Identification.}, a single-carrier waveform with quadrature phase-shift keying modulation is generated. Then, it is oversampled by a factor of 4, and impairments are added to the oversampled signal. In these simulations, only filter, PS, LO, and mixer components are identified based on $I/Q$ GI, QO, PN, FO, and $I/Q$ offset impairments. RF impairment models given in Section~\ref{Section2} are used to generate impairments. Also, the parameters of these impairments caused by the corresponding components are given in Table~\ref{Impairment_parameters}. Based on these range values, the impairments are randomly generated. Different threshold levels are determined for high, middle, and low-quality devices to determine the distorted components. If the impairment value is in the range interval, it is considered a distorted component, otherwise an undistorted component. Since the impairments are generated randomly and independently, a maximum of four different distorted or undistorted components can affect the generated signal. 60000 samples are generated for each scenario (high, middle, and low-quality devices), where 40000 are used for training while 10000 samples are used for both validation and testing. Keras, an open-source ML library under the Python environment, is used to implement the proposed DL algorithm for hardware problem identification. It should be noted that by training the network according to newly added hardware distortion, every type of distorted hardware in the RF system can be identified by making straightforward changes in the design.

\begin{table}[t]
\centering
\caption{RF impairments parameters.}
\resizebox{0.99\columnwidth}{!}{
\begin{tabular}{|c|c|c|ccc|}
\hline
\multirow{2}{*}{\textbf{Impairment}} & \multirow{2}{*}{\textbf{Range}} & \multirow{2}{*}{\textbf{Component}} & \multicolumn{3}{c|}{\textbf{Thresholds}}                                                                                                     \\ \cline{4-6} 
                                                      &                                                  &                            & \multicolumn{1}{c|}{\textbf{High-quality}}  & \multicolumn{1}{c|}{\textbf{Middle-quality}}     & \textbf{Low-quality}        \\ \hline
$I/Q$ GI                                 & 0 to 1                                           & Filter                     & \multicolumn{1}{c|}{\textgreater{}0.2}                    & \multicolumn{1}{c|}{\textgreater{}0.4}                    & \textgreater{}0.6                    \\ \hline QO                                     & -90 to 90                                        & PS             & \multicolumn{1}{c|}{\textless{}-20 or \textgreater{}20}   & \multicolumn{1}{c|}{\textless{}-40 or \textgreater{}40}   & \textless{}-60 or \textgreater{}60   \\ \hline
PN                                           & 0 to 90                                          & LO                         & \multicolumn{1}{c|}{\textgreater{}20}                     & \multicolumn{1}{c|}{\textgreater{}40}                     & \textgreater{}60                     \\ \hline
FO                                      & 0 to 100                                         & LO                         & \multicolumn{1}{c|}{\textgreater{}20}                     & \multicolumn{1}{c|}{\textgreater{}40}                     & \textgreater{}60                     \\ \hline
$I/Q$ offset                                          & -0.5 to 0.5                                      & Mixer                      & \multicolumn{1}{c|}{\textless{}-0.1 or \textgreater{}0.1} & \multicolumn{1}{c|}{\textless{}-0.2 or \textgreater{}0.2} & \textless{}-0.3 or \textgreater{}0.3 \\ \hline
\end{tabular}}
\label{Impairment_parameters}
\end{table}

\begin{table}[]
\centering
\caption{The performances of the proposed algorithm.}
\resizebox{0.99\columnwidth}{!}{
\begin{tabular}{|c|c|c|c|c|}
\hline
       & \textbf{$\alpha$}          & \textbf{$\pi$}            & \textbf{$\psi$}           & \textbf{$F_1$-score }       \\ \hline
\textbf{Filter} & 0.976-0.955-0.930 &0.974-0.956-0.928 &0.978-0.955-0.933 & 0.976-0.956-0.930 \\ \hline
\textbf{Mixer}  & 0.990-0.983-0.978 & 0.988-0.981-0.978 &0.992-0.986-0.976 & 0.990-0.983-0.977 \\ \hline
\textbf{LO}     & 0.997-0.994-0.990 &0.997-0.992-0.987&0.997-0.997-0.993 &0.997-0.994-0.990 \\ \hline
\textbf{PS}     &0.857-0.790-0.695 &0.847-0.793-0.697 & 0.871-0.792-0.702 &0.859-0.792-0.700 \\ \hline
\end{tabular}}
\label{Table_results}
\end{table}

\par Accuracy ($\alpha$), precision ($\pi$), recall ($\psi$), and $F_1$-score performance metrics are widely used to evaluate a classifier's performance. The accuracy metric is a ratio of true predicted observations to the total number of observations ($\alpha=\frac{\xi+\sigma}{\xi+\sigma+\upsilon+\mu}$), where $\xi$, $\sigma$, $\upsilon$, and $\mu$ signify the number of true positive, true negative, false positive, and false negative, respectively. The precision metric is the percentage of actually positive predicted to all observations predicted positive ($\pi=\frac{\xi}{\xi+\upsilon}$). The recall is the ratio of positive observations predicted correctly to the total number of observations in the actual class ($\psi=\frac{\xi}{\xi+\mu}$). Finally, $F_1$-score performance metric is used to measure the overall performance of the classifier algorithm by taking the harmonic average of precision and recall ($F_1$-score$=2\times \frac{\pi\times\psi}{\pi+\psi}$).

\par The simulation results are shown in Table~\ref{Table_results} for high, middle, and low-quality devices, respectively. This table shows that the proposed algorithm can identify all of the distorted hardware components. This is consistently true for all quality metrics. Moreover, the general performance is better when the high-quality device is used since it is easy to distinguish between distorted and non-distorted signals. On the other hand, the performance decreases when the quality of the device decreases. Furthermore, some of the distorted hardware components are identified with higher accuracy. This is because the effect of some distorted RF components is more easily distinguishable. For example, LO distortion can be easily identified since its effect is circularity in the constellation diagram as illustrated in Fig.~\ref{hardware_impairments} and this is an obvious distinguishable feature. Therefore, estimation difficulty is not uniform for all of the hardware distortions.

\par
\begin{figure}[!t]
\centering
\resizebox{0.99\columnwidth}{!}{
\begin{tabular}{cc}
\includegraphics{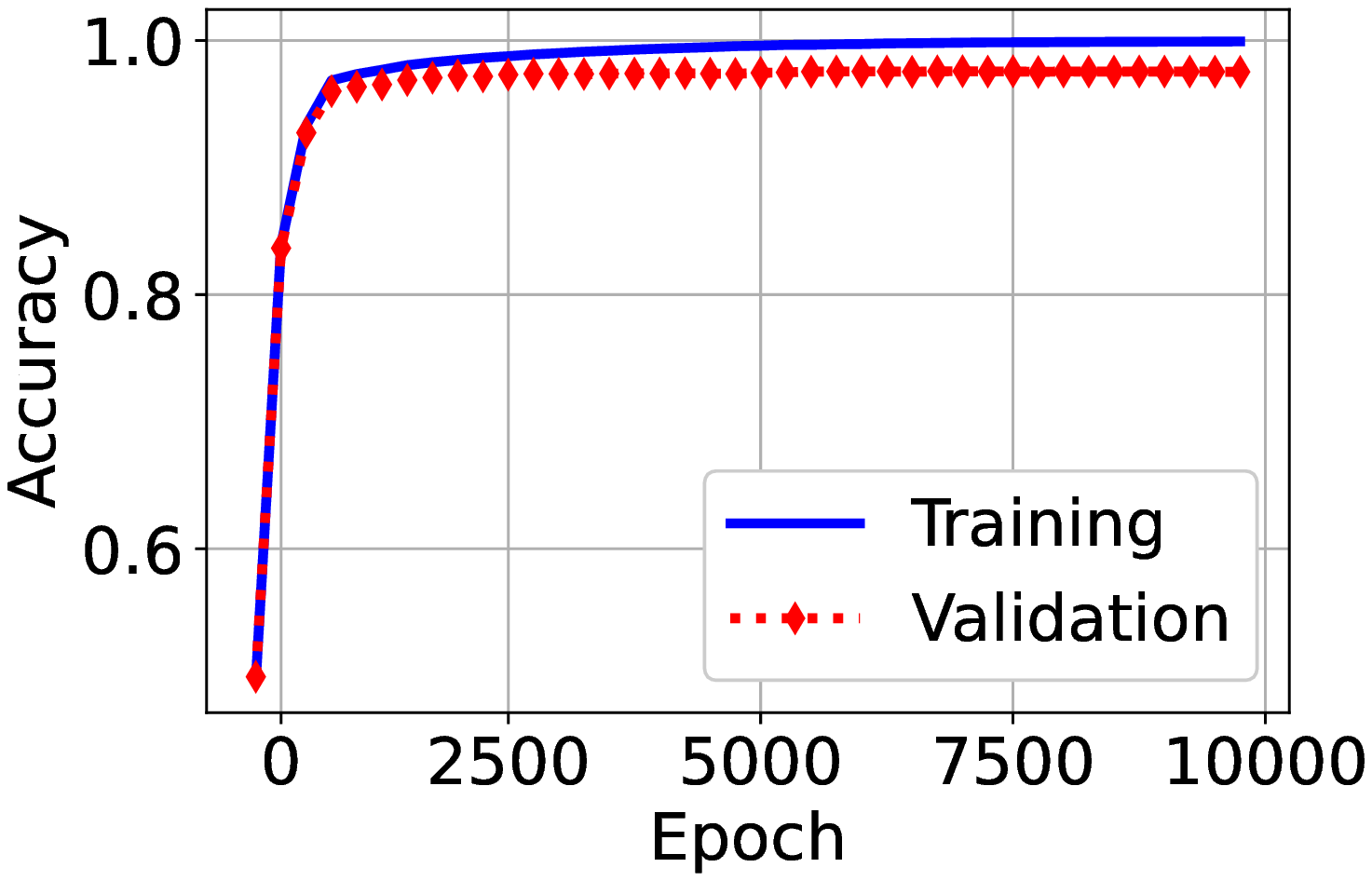}&
\includegraphics{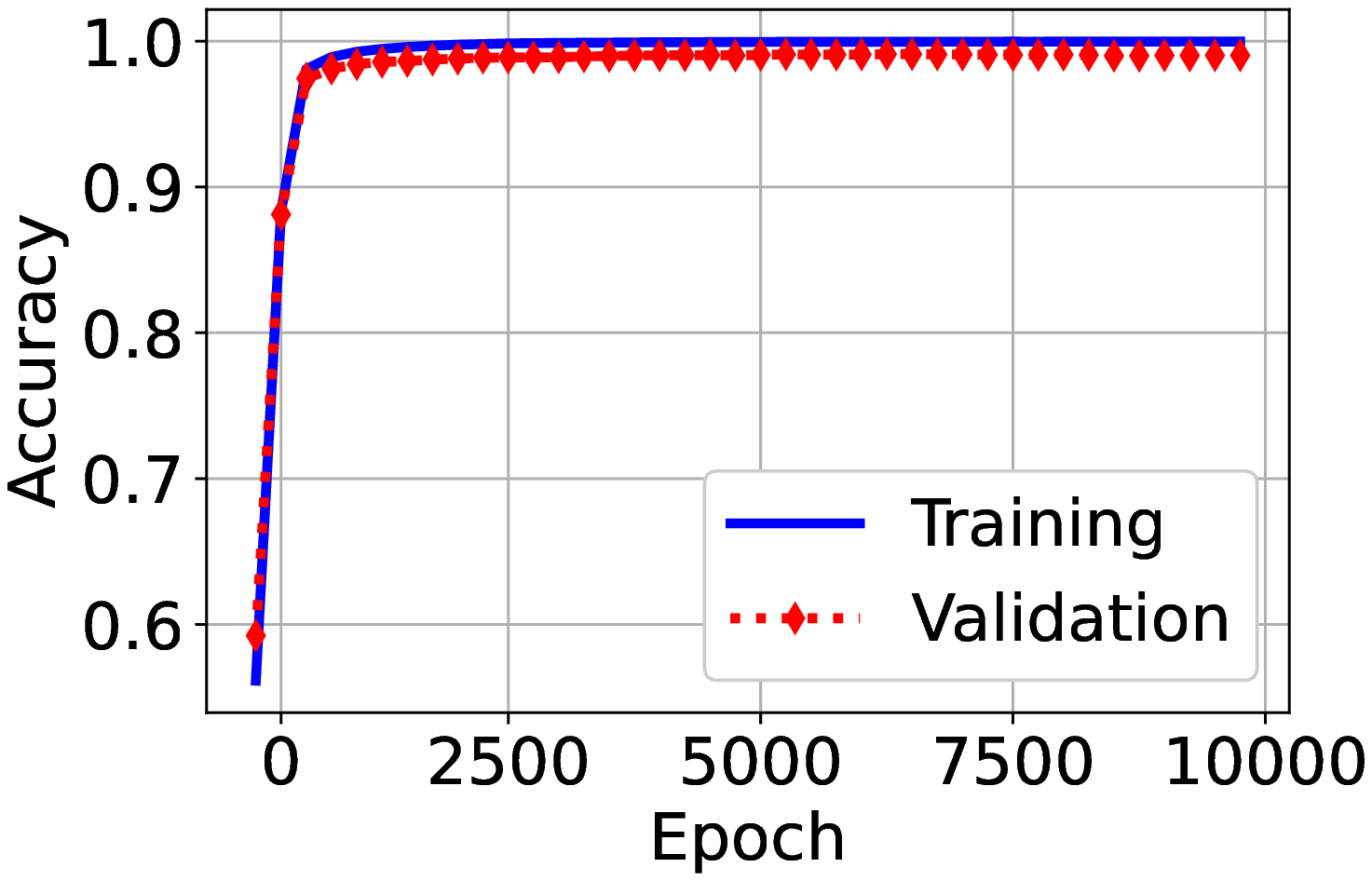} \\
\hspace{+2cm}\Huge{(a)} & \hspace{+2cm}\Huge{(b)}  \\
\includegraphics{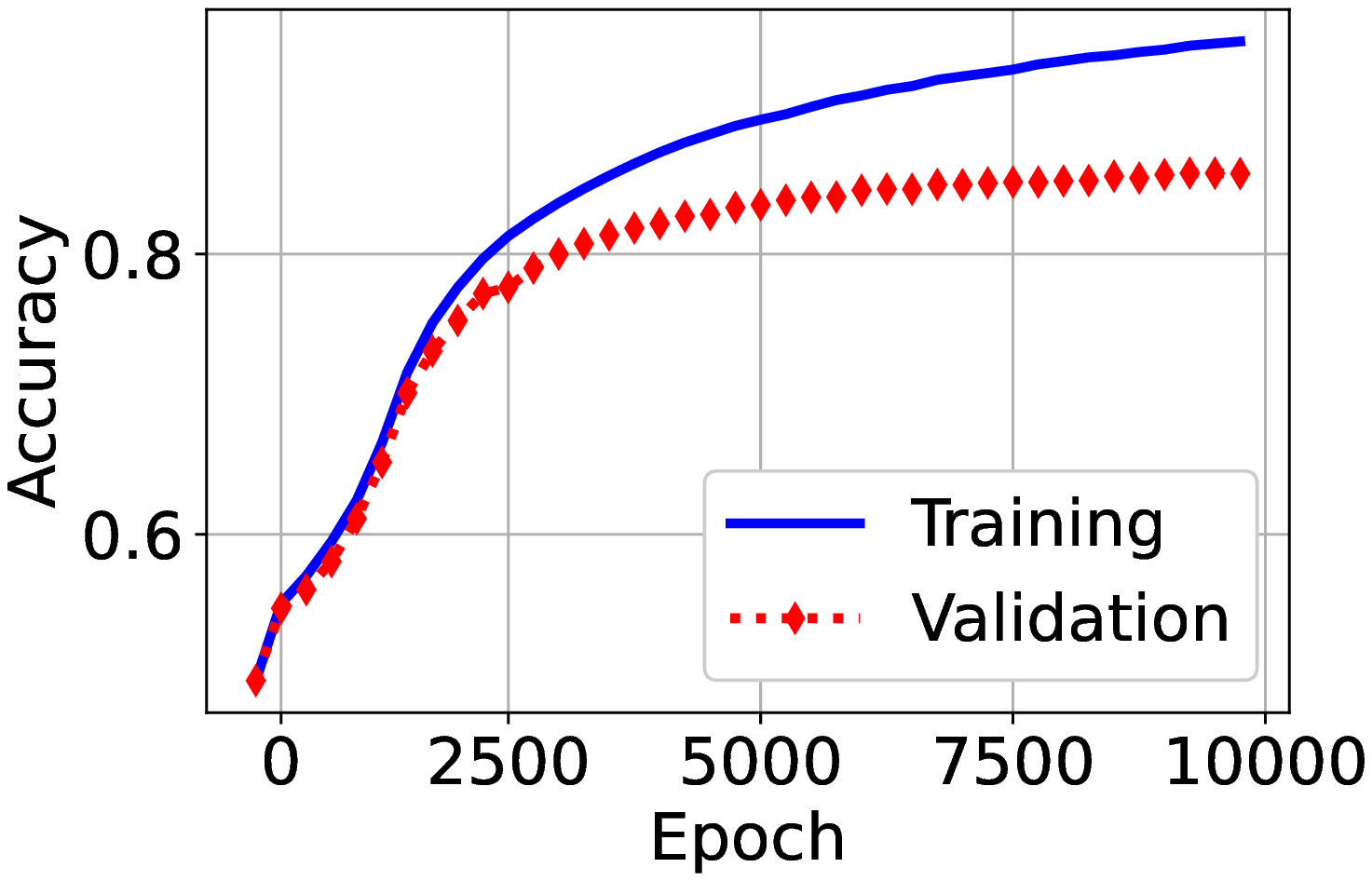}&
\includegraphics{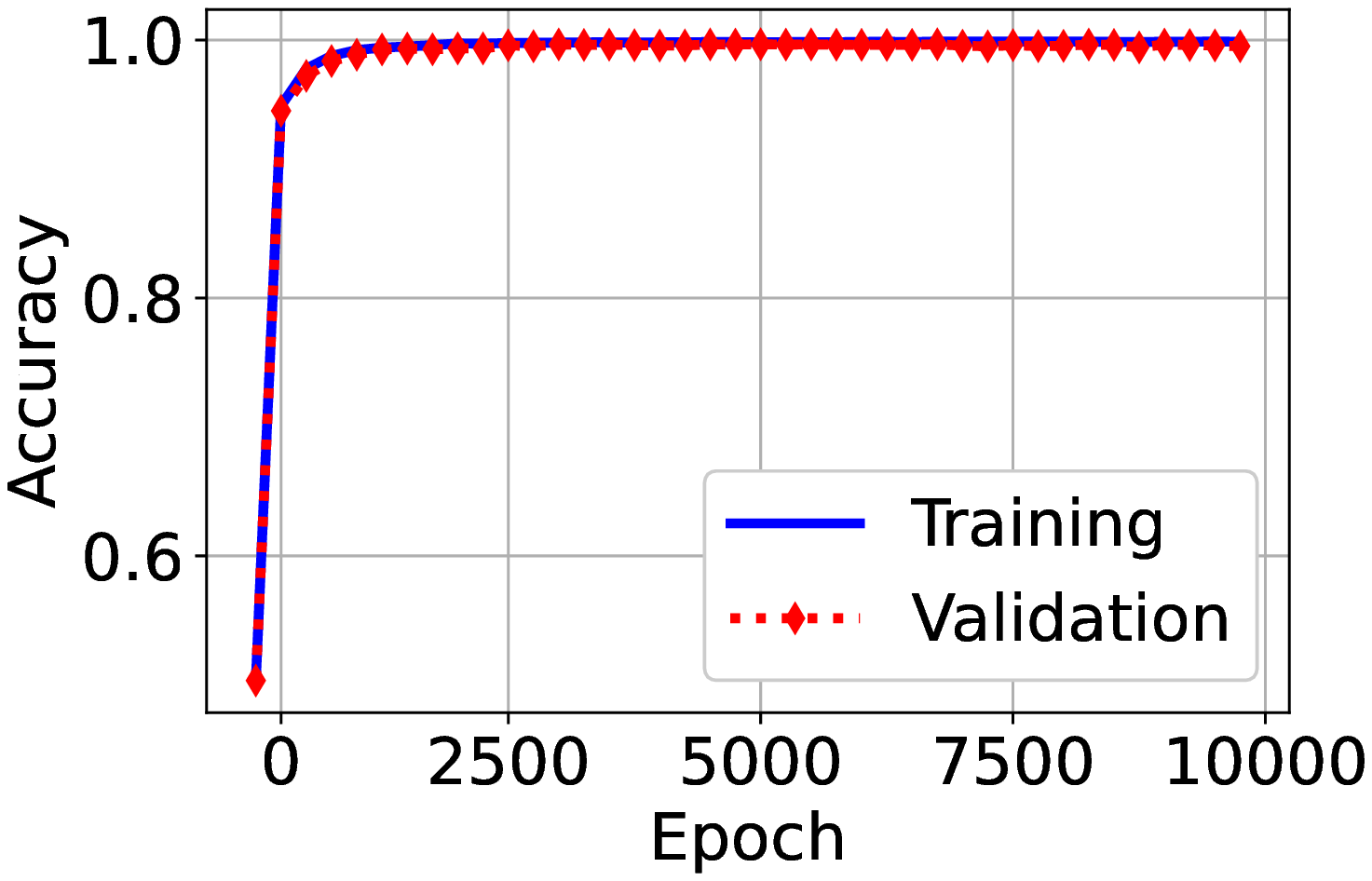}\\
\hspace{+2cm}\Huge{(c)} & \hspace{+2cm}\Huge{(d)} 
\end{tabular}}
\caption{Accuracy graphs for (a) filter, (b) mixer, (c) PS, and (d) LO.}
\label{accuracy} 
\end{figure}

 The generalization capability of a DL algorithm is also an essential success criterion. From an ML perspective, the training samples should not be memorized by a trained algorithm. The identification accuracies of the distorted RF components in both training and validation parts are shown in Fig.~\ref{accuracy}. The figure clearly shows that the accuracy of both training and validation sets converges. This validates the generalizability of the proposed algorithm, as neither overfitting nor underfitting is observed.

\section{Conclusions}
\label{Section5}

\par This paper highlighted the importance of hardware problem identification and showed the difficulty of the identification problem when multiple distorted hardware components exist in the communication system. Then, the use of deep MTL for this problem was proposed, and a deep MTL-based model was designed. Simulation results showed that the proposed algorithm can identify the distorted components with high performance. This was validated by accuracy, precision, recall, and $F_1$-score performance metrics in different scenarios. Also, the simulations verified that neither overfitting nor underfitting was observed with the designed DL algorithm. Last but not least, the proposed algorithm works automatically to identify distorted components. In future work, the proposed algorithm will be investigated in the real environment. 

\bibliographystyle{IEEEtran}

\end{document}